# Full *ab initio* band structure analysis of interband and intraband contributions for third harmonic generation coefficient of bulk silicon: implementation and application of the sum-over-states method

You-Zhao Lan*[1]

*Institute of Physical Chemistry, College of Chemistry and Life Sciences, Zhejiang Normal University, Zhejiang, Jinhua 321004, China*

**Abstract**

We fully implement the Aversa and Sipe sum-over-states formulism and make a full *ab initio* band structure analysis of interband and intraband contributions for the third-order nonlinear optical susceptibilities of bulk silicon. The band structure and momentum matrix elements were calculated by using the highly accurate all-electron full potential linearized augmented plane wave method within the local density approximation. The convergence tests including the scissor correction with different k-points meshes and empty states were performed. Both real and imaginary parts of susceptibility were directly calculated and checked by the Kramers–Kronig relation. The converged results are compared with other theoretical and experimental ones and in agreement with the recent *ab initio* real-time-based calculation. The nonlinear optical coefficient comes from three parts: the pure interband contribution (Pinter), the modulation of interband terms by intraband terms (Pmod), and the intraband contribution (Jintra). For each part, the origin of enhanced peaks is explored by tracing the sum-over-states process. The interband contribution is found to be dramatically modulated by the intraband contribution.

---
[1] Corresponding author: Youzhao Lan; Postal address: Institute of Physical Chemistry, College of Chemistry and Life Sciences, Zhejiang Normal University, Zhejiang, Jinhua 321004, China; Fax: +086 579 82282269; E-mail address: lyzhao@zjnu.cn

## 1. Introduction

In the last three decades, the perturbation-theory-based sum-over-states (SOS) method [1] has been widely used to calculate the optical polarizability of isolated atoms or molecules [2–4]. Using the SOS method, one can not only carry out the frequency-dependent optical response calculation but also investigate the electronic origin of the optical response. The former produces the theoretical results that more directly compared to experiments, as the nonlinear optical measurements are performed at different optical frequencies. The latter is helpful in identifying which excited states play a significant role in the optical response, then analyzing the charge transition contribution to these selective excited states, and ultimately identifying which functional groups dominate the optical response of the whole molecule. The corresponding findings will guide us to design and synthesize the material with a large nonlinear optical response [2,3].

For condensed semiconductor materials, the SOS method has also been developed to determine their linear and nonlinear optical susceptibilities [5–11]. For the linear optical susceptibility $\chi^{(1)}$, we can easily implement the SOS calculations providing that the band structure and momentum matrix elements are obtained. For the nonlinear optical susceptibilities such as $\chi^{(2)}$ and $\chi^{(3)}$, however, the difficulty increases rapidly owing to the complexity of the equations [6,7,9–15]. For $\chi^{(2)}$, the theoretic technique has been developed to a rather sophisticated level for both static and dynamic cases. At the outset, the static and dynamic calculations were separately developed because merging the static ($\omega \to 0$) calculation into the dynamic one was hindered by apparently diverging terms in the SOS equations [13,14,16,17] (i.e., factors of $\omega^{-1}$ and $\omega^{-2}$). To overcome the difficulty of unphysical divergences, Sipe and Ghahramani (SG) [10] developed the formalism for calculating the nonlinear optical coefficients within the independent-particle approximation. SG eliminated the unphysical divergences by a carefully separate treatment of interband and intraband motion and provided the detailed expressions for the calculation of second harmonic generation. These detailed expressions have been widely used to calculate the second harmonic generation coefficients of semiconductors [6–9,11,18–21]. These calculations provide much valuable information to understand the second order nonlinear optical response of semiconductors.

For $\chi^{(3)}$, application of the SOS technique is much less than for $\chi^{(2)}$ because of the complexity of equations. The connection between static and dynamic calculation was also plagued by the apparently diverging term [12,15]. SG has pointed that the divergence-free expressions of $\chi^{(3)}$ can be developed in a similar way to $\chi^{(2)}$ but the derivation will be a formidable task. On the other hand, also within the independent-particle approximation, by applying the perturbation theory to the dynamical equation of the electronic density operator and using a so-called length-gauge formulation, Aversa and Sipe (AS) [5] presented well-behaved, general expressions for $\chi^{(2)}$ and $\chi^{(3)}$ for arbitrary frequency mixings in a simpler way than SG. The expressions for $\chi^{(2)}$ are in well consistent with those shown in SG. Significantly, the derivation of expressions for $\chi^{(3)}$ from AS requires much fewer efforts than from SG. The expressions for $\chi^{(3)}$ were given explicitly in the divergence-free form. Clearly, these expressions are also helpful in understanding the third-order nonlinear optical response of semiconductors. However, to our best knowledge there has been no implementation of $\chi^{(3)}$ based on these expressions so far. Although there are a few reports [12,15,22–25] about calculations of $\chi^{(3)}$ based on the SOS method, these works are based on the two-band or empirical tight-banding and semi-*ab initio* band models. The more accurate band model calculations are required for the calculation of $\chi^{(3)}$ which is highly sensitive to the details of the band structures and wave functions [15].

The purpose of this work is two-fold. One is to fully implement the AS SOS formulism [5] to calculate the $\chi^{(3)}$. The other is to perform a full *ab initio* band structure analysis of interband and intraband contributions for third-order nonlinear optical susceptibilities of bulk silicon. The band structure is calculated by using the highly accurate all-electron full potential linearized augmented plane wave (FP-LAPW) method [26–28] within the local density approximation. The calculated susceptibilities are in agreement with the recent *ab initio* real-time-based computational approach combined with the Berry-phase formulation of the dynamical polarization [29]. By tracing the SOS process, we easily extract the transition term with a significant contribution to nonlinear coefficients and identify the $\omega$, $2\omega$, or $3\omega$ resonant contributions to the peak in the frequency-dependent nonlinear optical spectra.

In Sec. 2, we outline the formalism of AS and point out several critical points of implementation. In Sec. 3, the computational details for an application to silicon bulk are given. In Sec. 4, we discuss the scissor correction, convergence tests with k-points meshes and empty states, and Kramers–Kronig relation tests of $\chi^{(3)}(\omega)$, and compare our results with other theoretical ones as well as experiments, and analyze the origin of interband and intraband contributions to $\chi^{(3)}(\omega)$. Finally, we summarize our results in Sec. 5.

## 2. Formulism and implementation

The equations used to calculate the third-order response function were originally obtained by AS [5] who used the length-gauge formalism based on the position operator **r** (Ref.[30]). For convenience of reading, we inherit the AS's notations. The third-order susceptibility tensor represented by $\chi^{(3)}_{dcba}(-\omega_3;\omega_\gamma,\omega_\beta,\omega_\alpha)$, where $\omega_3=\omega_\gamma+\omega_\beta+\omega_\alpha$, is decomposed to $\chi^{(3)}_\chi$ and $\chi^{(3)}_\sigma$ based on the decomposition of the physical contributions to the polarization [5], namely, $d\mathbf{P}/dt = d\mathbf{P}_\chi/dt + \mathbf{J}_\sigma$. The final expressions for $\chi^{(3)}_\chi$ and $\chi^{(3)}_\sigma$ are as follows:

$$\frac{\chi^{(3)}_{\chi,dcba}}{C} = \sum_{l,m,n,p,k} \frac{r^d_{mn}}{\omega_{nm}-\omega_3}\left[\frac{r^c_{nl}}{\omega_{lm}-\omega_2}\left(\frac{r^b_{lp}r^a_{pm}f_{mp}}{\omega_{pm}-\omega_1} - \frac{r^a_{lp}r^b_{pm}f_{pl}}{\omega_{lp}-\omega_1}\right) - \left(\frac{r^b_{nl}r^a_{lp}f_{pl}}{\omega_{lp}-\omega_1} - \frac{r^a_{nl}r^b_{lp}f_{ln}}{\omega_{nl}-\omega_1}\right)\frac{r^c_{pm}}{\omega_{np}-\omega_2}\right]$$

$$+ i\sum_{l,m,n,k} \frac{r^d_{mn}}{\omega_{nm}-\omega_3}\left[\frac{1}{\omega_{nm}-\omega_2}\left(\frac{r^b_{nl}r^a_{lm}f_{ml}}{\omega_{lm}-\omega_1} - \frac{r^a_{nl}r^b_{lm}f_{ln}}{\omega_{nl}-\omega_1}\right)\right]_{;k^c} \quad (1)$$

$$+ i\sum_{l,m,n,k} \frac{r^d_{mn}}{\omega_{nm}-\omega_3}\left[\frac{r^c_{nl}}{\omega_{lm}-\omega_2}\left(\frac{r^a_{lm}f_{ml}}{\omega_{lm}-\omega_1}\right)_{;k^b} - \left(\frac{r^a_{nl}f_{ln}}{\omega_{nl}-\omega_1}\right)_{;k^b}\frac{r^c_{lm}}{\omega_{nl}-\omega_2}\right]$$

$$- \sum_{m,n,k} \frac{r^d_{mn}}{\omega_{nm}-\omega_3}\left[\frac{1}{\omega_{nm}-\omega_2}\left(\frac{r^a_{nm}f_{mn}}{\omega_{nm}-\omega_1}\right)_{;k^b}\right]_{;k^c}$$

and

$$\frac{\chi^{(3)}_{\sigma,dcba}}{C} = \sum_{mnk}\frac{1}{\omega_{mn}-\omega_2}\left(\frac{r^c_{nm}}{\omega_{mn}+\omega_3-\omega_2}\right)_{;k^d}\left(\frac{r^a_{mn}f_{nm}}{\omega_{mn}-\omega_1}\right)_{;k^b}$$

$$+ i\sum_{lmnk}\frac{r^a_{mn}r^b_{lm}}{\omega_{ln}-\omega_2}\left(\frac{f_n}{\omega_{mn}-\omega_1} + \frac{f_l}{\omega_{lm}-\omega_2+\omega_1}\right)\left(\frac{r^c_{nl}}{\omega_{nl}-\omega_3+\omega_2}\right)_{;k^d} \quad (2)$$

where $C = e^4K/\hbar^3$ with the $K$ factor depending on the particular combination [1] of $\omega_\gamma$, $\omega_\beta$, and $\omega_\alpha$, for example, $K$ is 1/4 for the third harmonic generation (THG) polarizability $\chi^{(3)}_{dcba}(-3\omega;\omega,\omega,\omega)$, $\omega_1$ and $\omega_2$ are defined by $\omega_1 = \omega_\alpha$ and $\omega_2 = \omega_\alpha + \omega_\beta$, respectively. $\omega_{mn} = \omega_m - \omega_n$ is the energy difference between the bands $m$ and $n$, $f_{mn} = f_m - f_n$ is the difference of the Fermi distribution functions, the indices of $a$, $b$, and $c$ are Cartesian directions, and all four band indices $l$, $m$, $n$, $p$ are different (one exception is shown below) because $r_{mn}$ [$= p_{mn}/(im\omega_{mn})$] is defined to be zero unless $n \neq m$ (Ref.[5]). So, to calculate the $\chi^{(3)}_{dcba}$ by using Eqs.1 and 2, we first have to obtain the band structure of periodic system and the momentum matrix elements.

As mentioned by AS, Eqs.1 and 2 remain to be symmetrized to satisfy the intrinsic permutation symmetry, that is, invariance under the possible permutations of $(a, \omega_\alpha)$, $(b, \omega_\beta)$, and $(c, \omega_\gamma)$. Here, we focus on the THG polarizability $\chi^{(3)}_{dcba}(-3\omega;\omega,\omega,\omega)$ whose intrinsic permutation symmetry can be readily performed by

$$\chi^{(3)}_{dcba}(-3\omega;\omega,\omega,\omega) = \frac{1}{6}\left[\begin{array}{l}\chi^{(3)}_{dcba}(-3\omega;\omega,\omega,\omega) + \chi^{(3)}_{dcab}(-3\omega;\omega,\omega,\omega) + \chi^{(3)}_{dbca}(-3\omega;\omega,\omega,\omega) \\ + \chi^{(3)}_{dbac}(-3\omega;\omega,\omega,\omega) + \chi^{(3)}_{dabc}(-3\omega;\omega,\omega,\omega) + \chi^{(3)}_{dacb}(-3\omega;\omega,\omega,\omega)\end{array}\right] \quad (3)$$

During the implementation of Eqs.1 and 2, the frequency $\omega$ should be understood as $\omega + i\eta$ with a small positive $\eta$ value and the following expressions were used,

$$\Delta^a_{nm} \equiv (\omega_{nm})_{;k^a} = (p^a_{nn} - p^a_{mm})/m \tag{4}$$

$$\left(\frac{r^a_{nm}}{\omega_{nm}-\omega}\right)_{;k^b} = \frac{r^a_{nm;k^b}}{\omega_{nm}-\omega} - \frac{r^a_{nm}\Delta^b_{nm}}{(\omega_{nm}-\omega)^2} \tag{5}$$

$$r^b_{nm;k^a} = \frac{r^a_{nm}\Delta^b_{mn} + r^b_{nm}\Delta^a_{mn}}{\omega_{nm}} + \frac{i}{\omega_{nm}}\sum_{l \neq n,m}\left(\omega_{lm}r^a_{nl}r^b_{lm} - \omega_{nl}r^b_{nl}r^a_{lm}\right) \tag{6}$$

$$\Delta^b_{nm;k^a} = \sum_l\left[\omega_{lm}\left(r^a_{ml}r^b_{lm} + r^b_{ml}r^a_{lm}\right) - \omega_{ln}\left(r^a_{nl}r^b_{ln} + r^b_{nl}r^a_{ln}\right)\right] \tag{7}$$

where the ;**k** operator represents a generalized derivative introduced by AS and **p** is momentum operator.

There is no obvious divergence in Eqs.1 and 2 except that the first summation of Eq.1 shows an apparent divergence arising from both a lack of $r_{lm}$ and $r_{np}$ elements in numerators and a factor of $1/\omega_2$ for $\chi^{(3)}_{dcba}(0;0,0,0)$ when $l = m$ and $n = p$. Introducing intrinsic permutation symmetry and relabeling indices, we reduced these two troublesome terms (TTT) to

$$\chi^{(3),\text{TTT}}_{\chi,dcba} = C\sum_{l,m,n,k} f_{ml}\left[\frac{r^d_{nm}r^c_{mn}}{(\omega_{mn}-\omega_3)} - \frac{r^c_{nm}r^d_{mn}}{(\omega_{nm}-\omega_3)}\right]\left[\frac{r^b_{ml}r^a_{lm}}{(\omega_{lm}-\omega_1)(\omega_{lm}+\omega_2-\omega_1)}\right] \tag{8}$$

In evaluation of the matrix elements $r_{mn}$ as $p_{mn}/(im\omega_{mn})$, a numerical problem occurs when the bands $m$ and $n$ are nearly degenerate. As mentioned by SG, one can always choose the appropriate wave functions for the bands $m$ and $n$ such that the matrix elements $r_{mn}$ (or $p_{mn}$) vanishes. Therefore, for nearly degenerate bands $m$ and $n$ decided by a small cutoff value, e.g., $\omega_{mn} \leq 0.001$ a.u., we set $r_{mn}$ to be zero, in consistent with the definition of $r_{mn}$ above. This strategy was smoothly used by Rashkeev *et al.*[7] in their calculations of frequency-dependent second-order optical response of semiconductors.

**3. Computational details**

We applied our implementation to the calculation of THG of cubic silicon crystal (Si) with a lattice parameter of 5.43Å. The wave functions and momentum matrix elements were computed with the highly accurate all-electron FP-LAPW method [26–28] within the local density approximation as implemented in the ELK code [31]. Since the LDA calculation underestimates the band gap of Si, we applied the widely used scissor correction [32] in the optical calculation. Both the real and imaginary parts of $\chi^{(3)}$ were directly calculated and checked by the Kramers–Kronig relation (KKR) [33]. The maximum angular momentum used for APW functions is $l_{max} = 8$. Since the nonlinear optical calculation possibly requires much denser k-points mesh and more empty states than the linear optical calculation [6,9], we performed the convergence tests of $\chi^{(3)}$ on the 10×10×10, 20×20×20, 30×30×30, and 40×40×40 k-points meshes and the number of empty states (10, 14, and 18) per atom. In terms of the limit of cubic symmetry, we only calculated two nonzero independent elements of $\chi^{(3)}$, namely, $\chi^{(3)}_{1111}$ and $\chi^{(3)}_{1212}$.

**4. Applications**

*4.1 Scissor correction*

It is well known that the LDA underestimates the band gap in semiconductors. Since the denominators of Eqs.1 and 2 depend on the $1/\omega^4_{nm}$ like factors, the underestimation of the band gaps definitely leads to an error in calculation of $\chi^{(3)}$. The simple and effective way is to introduce a so-called scissor correction, in which the band energies are shifted by a factor of $\Delta\omega$ and the momentum matrix elements are corrected by $p_{mn} = p_{mn}(1+f_{mn}\Delta\omega/\omega_{nm})$ (Ref.[32]). The scissor correction is derived by adding to the LDA Hamiltonian the scissor operator that is an effective self-energy [34,35]. The $f_{mn}\Delta\omega/\omega_{nm}$ is considered as a nonlocal contribution to the matrix element [34,35], in consistent with a point of view of the nonlocal exact density functional [36]. This nonlocal correction may have a significant contribution to the matrix element, as shown by Nastol *et al.* [32]. Their estimation for GaAs had shown that the $\Delta\omega/\omega_{nm}$ value is about 4.4 for a lowest conduction band *n*, and a highest valence band *m* near the *Γ* point. As shown in calculations of $\chi^{(2)}$ of GaAs and GaP, the magnitude of $\chi^{(2)}$ was dramatically improved by applying the scissor correction [32]. Using the FP-LAPW/LDA method, we obtain for Si an indirect gap of 0.46 eV which is lower by 0.69 eV than the experimental value of 1.15 eV. In the following calculations of $\chi^{(3)}$ we used a scissor correction of 0.69 eV.

*4.2 Convergence test*

The calculations of $\chi^{(2)}$ [7,9] have shown that a denser k-points mesh is required for nonlinear than linear optical properties in the irreducible Brillouin zone (IBZ). We performed the convergence tests on the 10×10×10, 20×20×20, 30×30×30, and 40×40×40 *k*-points meshes corresponding to 47, 256, 752, and 1661 *k*-points in IBZ, respectively. Figure 1a shows the LDA absolute values of $\chi^{(3)}_{1111}(\omega)$ for different *k*-points meshes. For $\omega > 0.5$eV, the $\chi^{(3)}_{1111}(\omega)$ values present a good convergence for four k-points meshes and agree with recently calculated results [29] based on an *ab initio* real-time-based computational approach (see Section 4.4). However, for $\omega < 0.5$ eV, $\chi^{(3)}_{1111}(\omega)$ presents a poor convergence and its static values are separately $0.23 \times 10^{-10}$, $2.41 \times 10^{-10}$, $5.22 \times 10^{-10}$, and $6.43 \times 10^{-10}$ esu for four k-points meshes. To explore the poor convergence at $\omega > 0.5$ eV, we show in Fig. 1b and 1c the convergence tests on the real parts of $\chi^{(3)}_{1111\chi}(\omega)$ (Eq.1) and $\chi^{(3)}_{1111\sigma}(\omega)$ (Eq.2), and observe that a poor convergence of $\chi^{(3)}_{1111}(\omega)$ in the low applied frequency region should be attributed to the poor convergence of $\chi^{(3)}_{1111\sigma}(\omega)$. In section 4.5, we will show that the existence of $(f_n/\omega_{mn})$-like terms in $\chi^{(3)}_{1111\sigma}(\omega)$ leads to a possible resonance in the low applied frequencies. In addition to the *k*-points convergence test, we have also run the tests on the number of empty states included in the SOS calculations. Figure 1d shows a good convergence behavior for the number of empty states (*i.e.*, 10, 14, and 18) per atom. For $\chi^{(3)}_{1212}(\omega)$, we obtain a similar behavior for convergence tests. In the following section, we will use the 30×30×30 results to make further discussions.

*4.3 KKR test*

The KKR [33] describes a general connection between the real and imaginary parts of complex optical functions. In previous researches on nonlinear optical calculations [7,15,37], to reduce the computational efforts, one often only calculated the imaginary part of optical function and used the KKR to calculate its real part. As shown in Eqs.1 and 2, the strategy based on the KKR of $\chi^{(3)}(\omega)$ does not simplify the calculation because the full extract of the imaginary parts of $\chi^{(3)}(\omega)$ is also very complicated. So we directly calculated the real and imaginary parts of $\chi^{(3)}(\omega)$ and used the KKR to check the validity of results. For THG, the real and imaginary parts of $\chi^{(3)}(\omega)$ satisfy the following KKR [33],

$$\text{Re}\left\{\chi^{(3)}_{dcba}(-3\omega_1;\omega_1,\omega_1,\omega_1)\right\} = \frac{2}{\pi}\int_0^\infty \frac{\omega_2 \text{Im}\left\{\chi^{(3)}_{dcba}(-3\omega_2;\omega_2,\omega_2,\omega_2)\right\}}{\omega_2^2 - \omega_1^2} d\omega_2 \qquad (9)$$

As an example, we shows in Fig. 2 the dispersions of the directly calculated real and imaginary parts of $\chi^{(3)}_{1111}(\omega)$ and of the real part of $\chi^{(3)}_{1111}(\omega)$ based on the KKR calculation. It is clearly shown that the directly calculated results are in well consistent with the KKR ones, which supports our correct implementations of Eqs.1 and 2.

*4.4 Comparisons with other theoretical results and experiments*

To get a better evaluation for our results, we compare our results with other theoretical reports [15,29,38–40] as well as experiments [41,42]. In Fig. 3, we compare our results based on 30×30×30 k-points with those based on the tight-binding with either semi-*ab-initio* (STB) or empirical parameters (ETB) as well as based on an *ab initio* real-time-based computational approach combined with the Berry-phase (BP) formulation of the dynamical polarization. And for clarity, we list in table 1 the theoretical and experimental values of $\chi^{(3)}_{1111}(\omega)$ (×$10^{-10}$ esu) at $\omega = 0.0$ and 1.16 eV. Overall, there are apparent differences between different theoretical values and between theoretical and experimental values. In the static case, Jha and Blqembergen [39] obtained a negative $\chi^{(3)}_{1111}(0)$ value of $-0.25 \times 10^{-10}$ esu within the completely-localized-bond approximation on the basis of simple tetrahedral bonding orbitals. They obtained a negative $\chi^{(3)}_{1111}(0)$ value owing to the vanishment of the term including matrix elements between bonding-bonding or anti-bonding-anti-bonding states in the limitation of the center-of-inversion symmetry of Si. Since this negative theoretical value dramatically differs from available experimental values [41,42], using a tight-binding model with retaining more interactions between bonds, Arya and Jha [38] appeared to obtain largely improved results for $\chi^{(3)}_{1111}(0)$, however, these results are not so good as concluded by Arya and Jha because they used a different definition as pointed by Moss *et al.*(See. ref.24 of [15]). Significantly, by considering the intraband Franz-Keldysh effect to calculate $\chi^{(3)}(\omega)$, Vechten and Aspnes [40] obtained a largely improved value of $0.20 \times 10^{-10}$ esu in good agreement with the experimental value as shown in table 1, even though they included only two bands ($\Gamma_{25'}$ and $\Gamma_{15}$ of Si). Their work also implies an importance of intraband contributions especially for crystals with small energy gaps. As a further improvement, Moss *et al.* [15] used both an empirical tight-binding (ETB) and a semi-*ab-initio* band structure technique with standard perturbation theory to calculate the $\chi^{(3)}(\omega)$ and obtained the results with a better agreement with the static experimental value. Note that there is a large uncertainty (±60%) in the static experimental value (table 1), which leads to the results ranging from $0.096 \times 10^{-10}$ to $0.384 \times 10^{-10}$ esu.

For the dynamic case, as shown in table 1, all the theoretical values at $\omega = 1.16$ eV dramatically deviate from the experimental value. However, from Fig. 3, we observe that all the theoretical methods yield a similar dispersion with $\omega > 0.5$ eV and the dispersion presents a peak due to a possible $2\omega$ or $3\omega$ resonance near 1.3 eV. Our results present a very similar line shape to the BP ones except that a small energy shift is observed in the position of the peaks. This small shift is due to the used different scissor corrections (0.71 eV for us and 0.60 eV for BP [29]). It should be noted that there is also a large uncertainty (±50%) in this experimental value and the measurement is often performed relative to a reference sample. So a direct comparison between theoretical and experiment values is often difficult [2].

*4.5 Interband and Intraband contributions*

Figure 4 shows the dispersions of interband and intraband contributions of $\chi^{(3)}(\omega)$. In terms of the decomposition of position operator ($r = r_i + r_e$, $r_i$ and $r_e$ are intraband and interband parts of $r$) [5], Eq.1 related to the interband part $r_e$ is referred to as the interband contribution while Eq.2 related to the intraband part $r_i$ is the intraband contribution (Jintra in Fig.4). Note that the later three summations of Eq.1 contain diagonal momentum matrix elements ($p_{nn}$ of Eq.4). So the later three summations of Eq.1 is considered as a modulation of interband terms by intraband terms (Pmod in Fig.4), similar to the decomposition of $\chi^{(2)}(\omega)$ [6,8,9], and the first summation of Eq.1 that does not contains diagonal momentum matrix elements is considered as the pure interband contribution (Pinter in Fig.4).

Firstly, as shown in Fig.4, Pinter presents the enhancement at some applied frequencies. For example, the Re[$\chi^{(3),\text{Pinter}}_{1111}(\omega)$] presents two peaks at $\omega = 1.55$ and 2.16 eV. To understand the origin of these two peaks, we traced the SOS calculations in terms of Eq.1 for these two applied frequencies. Figure 5 shows the trace of summation over all 27000 (30×30×30) k-points for Re[$\chi^{(3),\text{Pinter}}_{1111}(\omega)$] with $\omega = 1.55$ and 2.16 eV. We observe that a great number of k-points almost have no contribution to Re[$\chi^{(3)}_{1111}(\omega)$]. In detail, we show the distribution of contribution for each k-points in summation as insets of Fig. 5, where only absolute contributions larger than 1.0 a.u. are given for clarity. From insets, we can see that a few k-points (labeled by $a$, $b$, $c1$, $c2$, $d1$, $d2$, $e1$, $e2$, $f1$, and $f2$) have a relative large contribution to Re[$\chi^{(3),\text{Pinter}}_{1111}(\omega)$]. To check whether these relative large contributions are attributed to the possible $\omega$, $2\omega$, or $3\omega$ resonance in Pinter of Eq.1, we separately traced the SOS calculations for these k-points. Table 2 lists the single-particle channels with significant contributions to each k-points. Eq.11 shows the calculating expression for Pinter obtained by rearranging the dummy indices.

$$\frac{\chi_\chi^{(3)\text{Pinter}}}{C} = \sum_{l,m,n,p,k} \frac{r_{mn}^d}{\omega_{nm}-\omega_3}\left[\frac{r_{nl}^c}{\omega_{lm}-\omega_2}\left(\frac{r_{lp}^b r_{pm}^a f_{mp}}{\omega_{pm}-\omega_1} - \frac{r_{lp}^a r_{pm}^b f_{pl}}{\omega_{lp}-\omega_1}\right) - \left(\frac{r_{nl}^b r_{lp}^a f_{pl}}{\omega_{lp}-\omega_1} - \frac{r_{nl}^a r_{lp}^b f_{ln}}{\omega_{nl}-\omega_1}\right)\frac{r_{pm}^c}{\omega_{np}-\omega_2}\right]$$

$$= f_{mn} \sum_{l,m,n,p,k}\left[\frac{r_{mp}^d}{\omega_{pm}-\omega_3}\frac{r_{pl}^c}{\omega_{lm}-\omega_2}\frac{r_{ln}^b r_{nm}^a}{\omega_{nm}-\omega_1} - \frac{r_{pl}^d}{\omega_{lp}-\omega_3}\frac{r_{ln}^c}{\omega_{np}-\omega_2}\frac{r_{nm}^a r_{mp}^b}{\omega_{nm}-\omega_1}\right.$$
$$\left. - \frac{r_{pl}^d}{\omega_{lp}-\omega_3}\frac{r_{mp}^c}{\omega_{lm}-\omega_2}\frac{r_{ln}^b r_{nm}^a}{\omega_{nm}-\omega_1} + \frac{r_{ln}^d}{\omega_{nl}-\omega_3}\frac{r_{pl}^c}{\omega_{np}-\omega_2}\frac{r_{nm}^a r_{mp}^b}{\omega_{nm}-\omega_1}\right] \quad (10)$$

Combining Eq.10 and these single-particle channels (table 2), we may identify which resonance leads to a significant contribution. For $\omega = 1.55$ eV (0.056 a.u.), the channel of $2(0.1920) \rightarrow 7(0.2946) \rightarrow 8(0.3187) \rightarrow 4(0.1940)$ with 70% contribution to the final result corresponds to a term of $(m = 2) \rightarrow (n = 7) \rightarrow (l = 8) \rightarrow (p = 4)$ in summation. Inspecting the denominators of Eq.10, we can see that the $2\omega$ resonance leads to a significant contribution because both $\omega_{lm}$ (= 0.1267 a.u.) and $\omega_{np}$ (= 0.1006 a.u.) with a scissor correction of 0.026 a.u. are close to $\omega_2$ (= $2\omega = 2 \times 0.056 = 0.112$ a.u.). A similar analysis for $\omega = 2.16$ eV (0.079 a.u.) also shows that the $2\omega$ resonance leads to a significant contribution because both $\omega_{lm}$ (= 0.1483 a.u.) and $\omega_{np}$ (= 0.1561 a.u.) with a scissor correction of 0.026 a.u. are close to $\omega_2$ (= $2\omega = 2 \times 0.079 = 0.158$ a.u.). For other peaks shown in Fig.5, we can perform similar analyses to check the origin of nonlinear optical response.

Secondly, Pmod has the same magnitude of contribution as Jintra at $\omega < 0.5$ eV and presents similar enhancements to Pinter. For example, at $\omega = 1.63$ eV, $\text{Re}[\chi^{(3),\text{Pmod}}_{1111}(\omega)]$ has a positive contribution (30556.8 a.u. = $1.04 \times 10^{-10}$ esu) to $\text{Re}[\chi^{(3)}_{1111}(\omega)]$. Similarly, we traced the SOS process of k-points which have significant contributions to $\text{Re}[\chi^{(3),\text{Pmod}}_{1111}(\omega)]$ in summation. Three values (173.2, 167.1, and 166.5 a.u.) are traced in the SOS process. These three largest values are dominated by the single-particle channels formed by bands [2(0.1922)$m$, 8(0.3185)$n$, and 4(0.1941)$l$] and [2(0.1921)$m$, 7(0.2945)$n$, and 4(0.1941)$l$]. In each channel, the existence of two $2\omega$ resonances leads to a significant contribution to $\text{Re}[\chi^{(3),\text{Pmod}}_{1111}(\omega)]$ because, for example, both $\omega_{82}$ (= 0.1263 a.u.) and $\omega_{84}$ (= 0.1244 a.u.) with a scissor correction of 0.026 a.u. are close to $\omega_2$ (= $2\omega = 2 \times 0.060 = 0.120$ a.u.). A similar case occurs for the channel formed by bands 2, 7, and 4. Additionally, the modulations of intraband term such as $\Delta_{nl}/(\omega_{nl} - \omega_2)^2$ also lead to Pmod with a large contribution because the small $(\omega_{nl} - \omega_2)^2$ term further magnifies the contribution from the $2\omega$ resonance.

Thirdly, Jintra shows a resonance to $\chi^{(3)}(\omega)$ in the low applied frequency region. In the static case, the real parts of $\chi^{(3)}_{1111}(0)$ for Pinter, Pmod, and Jintra contributions are 0.028, −0.216, and $5.41 \times 10^{-10}$ esu, respectively. According to Eqs.1 and 2, in the static case, the size of contribution is mainly determined by the energy difference between bands (*i.e.*, $\omega_{mn}$). In detail, Fig.6 shows the trace of the SOS process for the static case. Only parts of 27000 k-points dramatically contribute to the Re[$\chi^{(3),Jintra}_{1111}(0)$]. More clearly, we also show the distribution of contribution for each k-points in summation as an inset of Fig.6. As examples, we traced the SOS process of two largest values (labeled by *a* and *b* in the inset of Fig.6). These two largest values are dominated by the single-particle channels formed by bands 2(0.1796 a.u.), 3(0.1905 a.u.), and 4(0.1933 a.u.). We can see that small energy differences among these three bands (i.e., $\omega_{23} = -0.0109$ a.u., $\omega_{24} = -0.0137$ a.u., and $\omega_{34} = -0.0028$ a.u.) possibly lead to a small denominator, then to a large contribution to Re[$\chi^{(3),Jintra}_{1111}(0)$]. On the other hand, a small energy difference has not led to similar resonance for Pinter and Pmix in the low applied frequency region, as shown in Fig.4. A possible reason is that the Eq.2 for Jintra has $(f_{mn}/\omega_{mn})$ and $(f_n/\omega_{mn})$-like terms while the Eq.1 for Pinter and Pmod only have $(f_{mn}/\omega_{mn})$-like term. Figure 7 shows the dispersion behaviors of the first and second summations of Re[$\chi^{(3),Jintra}_{1111}(\omega)$]. We can see from Fig.7 that a resonance clearly comes from the second summation in the low frequencies. Furthermore, as shown in table 2, all the bands of *m* and *n* with a small energy difference $\omega_{mn}$ belong to the same band region (valance or conduction band), which leads to a zero value of $f_{mn}$, then to a zero contribution of $f_{mn}/\omega_{mn}$. However, in the valance band, the $(f_n/\omega_{mn})$-like term still alive because of $f_n = 1$, which leads to a large $f_n/\omega_{mn}$ value. So, the single-particle channels formed by bands 2(0.1796 a.u.), 3(0.1905 a.u.), and 4(0.1933 a.u.) will have a large contribution to Re[$\chi^{(3),Jintra}_{1111}(0)$] owing to the $(f_3/\omega_{34})$ or

($f_4/\omega_{34}$) terms.

## 5. Conclusions

We have fully implemented the Aversa and Sipe sum-over-states formulism and performed a full *ab initio* band calculation of the frequency-dependent third harmonic generation of bulk silicon. The interband contribution will be dramatically modulated by the intraband contribution in the high order optical response. By tracing the SOS process, we clearly recognize the origin of peaks ($\omega$, $2\omega$, or $3\omega$ resonance) in both interband and intraband contributions. The $2\omega$ resonance seems to be a main reason for nonlinear enhancement of bulk silicon. Undoubtedly, The SOS method is preferable for understanding the mechanics of nonlinear optical response of semiconductor.

**Acknowledgements**

We appreciate the financial supports from Natural Science Foundation of China Project 21303164 and the computational supports from Zhejiang Key Laboratory for Reactive Chemistry on Solid Surfaces.

List of figure captions:

Figure 1. Convergence tests on $\chi^{(3)}(\omega)$ with (a, b, c) the 10×10×10, 20×20×20, 30×30×30, and 40×40×40 k-points meshes and (d) with the number of empty states per atom, *i.e.*, 10, 14, and 18 empty states per atom.

Figure 2. KKR test from imaginary (Im) to real (Re) part of $\chi^{(3)}(\omega)$. The dot and solid lines indicate the directly calculated results while the hollow square indicates the results based on the KKR calculation.

Figure 3. Dispersion of the two independent components of $\chi^{(3)}(\omega)$ based on our strategy, the tight-binding calculations with either semi-*ab initio* (STB) or empirical parameters (ETB), and *ab-initio* approach by means of the dynamical Berry phase (BP).

Figure 4. Dispersions of the pure interband (Pinter), a modulation of interband terms by intraband terms (Pmod), and intraband (Jinter) contributions of $\chi^{(3)}(\omega)$. P and J indicate $\mathbf{P}_\chi$ and $\mathbf{J}_\sigma$ components, respectively. For clarity, the pure interband contribution is increased by an order of magnitude (*i.e.*, ×10).

Figure 5. Traces of summation over all 27000 (30×30×30) k-points for the real part of $\chi^{(3),\text{Pinter}}_{1111}(\omega)$ with $\omega$ = 1.55 and 2.16 eV. Insets are the distribution of contribution (a.u.) per k-points in summation and for clarity only absolute contributions larger than 1.0 a.u. are shown.

Figure 6. Traces of summation over all 27000 (30×30×30) k-points for the real part of $\chi^{(3),\text{Jintra}}_{1111}(0)$. The inset is the distribution of contribution (a.u.) per k-points in summation and for clarity only absolute contributions larger than 500 a.u. are shown.

Figure 7. Dispersion behaviors of the first (Jintra-sum1) and second (Jintra-sum2) summations of $\chi^{(3),\text{Jintra}}_{1111}(\omega)$ (Eq. 2 in text).

Table 1. Theoretical and experimental values of $\chi^{(3)}_{1111}(\omega)$ ($\times 10^{-10}$ esu) at $\omega = 0.0$ and 1.16 eV.

|  | Methods | | | | |
|---|---|---|---|---|---|
|  | This work | STB | ETB | Other theoretic results | Expt. |
| $\omega = 0.0$ eV | 5.22 | 0.08 | 0.48 | –0.25 [a], 0.026 [b], 0.20 [c] | 0.24 ± 60% [e] |
| $\omega = 1.16$ eV | 0.83 | 1.3 | 2.2 | 0.84 [d] | 16.8 ± 50% [f] |

[a] Reference 39

[b] Reference 38

[c] Reference 40

[d] BP 29

[e] Reference 41,42

[f] This is a relative value measured relative to LiF at the same applied field ($\omega = 1.16$ eV).

Table 2. Selected contributions labeled in insets of Fig. 6 to Re[$\chi^{(3),Pinter}_{1111}(\omega)$] (a.u.). The single-particle channels with significant contributions to each k-points are given. 2(0.1920) indicates the band 2 with an eigenvalue of 0.1920 a.u. Note that for Si, since the primitive cell includes two silicon atoms and eight valence electrons were considered in calculations, bands 1, 2, 3, and 4 are the valance bands and the rest are conduction bands.

| Labels | Contributions [Percentages, Channels (Eigenvalues)] | Final results |
|---|---|---|
| $\omega$ = 1.55 eV (0.056 a.u.) | | |
| a | –22.52 [70%, 2(0.1920)→7(0.2946)→8(0.3187)→4(0.1940)] | –32.11 |
|   | –9.21 [29%, 4(0.1940)→8(0.3187)→7(0.2946)→2(0.1920)] |   |
| b | –23.02 [70%, 2(0.1922)→7(0.2945)→8(0.3185)→4(0.1941)] | –32.75 |
|   | –9.35 [29%, 4(0.1941)→8(0.3185)→7(0.2945)→2(0.1922)] |   |
| $\omega$ = 2.16 eV (0.079 a.u.) | | |
| c1 and c2 | 17.31 [42%, 3(0.0839)→6(0.2400)→5(0.2322)→4(0.0859)] | 41.36 |
|   | 23.80 [57%, 4(0.0859)→5(0.2322)→6(0.2400)→3(0.0839)] |   |
| d1 and d2 | 17.82 [42%, 3(0.0840)→6(0.2398)→5(0.2321)→4(0.0859)] | 42.51 |
|   | 24.42 [57%, 4(0.0859)→5(0.2321)→6(0.2398)→3(0.0840)] |   |
| e1 and e2 | 17.58 [42%, 3(0.0841)→6(0.2398)→5(0.2319)→4(0.0860)] | 42.01 |
|   | 24.17 [57%, 4(0.0860)→5(0.2319)→6(0.2398)→3(0.0841)] |   |
| f1 and f2 | 17.08 [42%, 3(0.0840)→6(0.2400)→5(0.2321)→4(0.0860)] | 40.88 |
|   | 23.56 [57%, 4(0.0860)→5(0.2321)→6(0.2400)→3(0.0840)] |   |

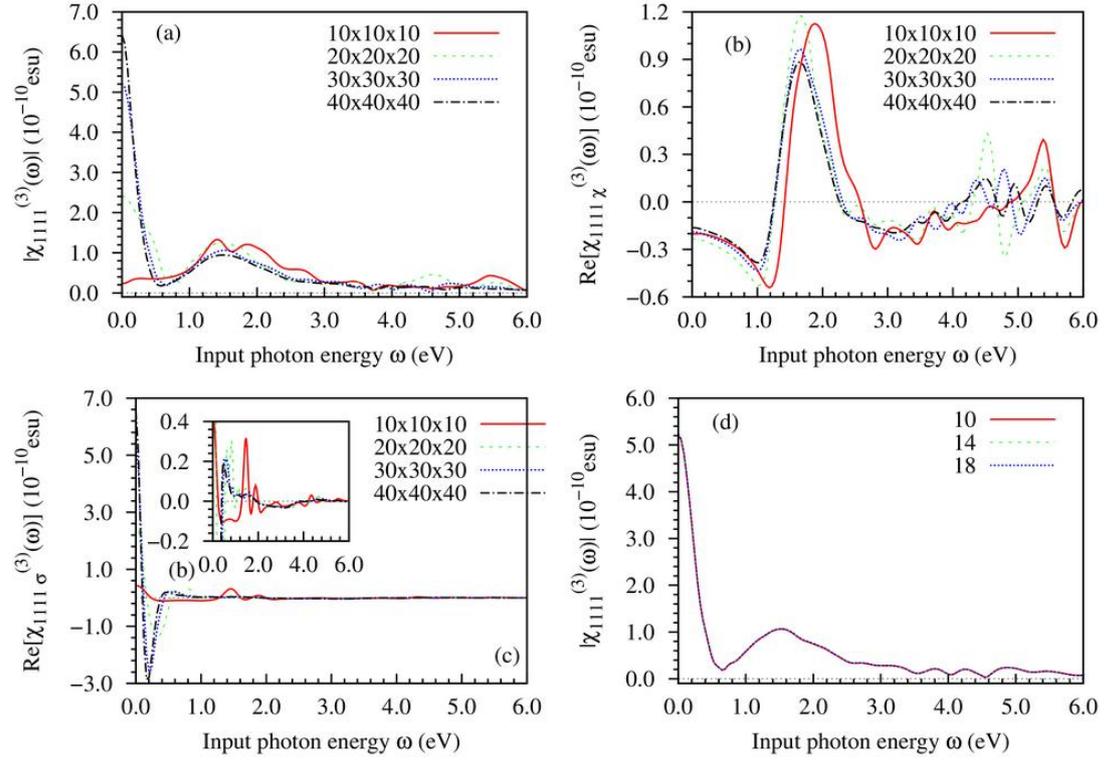

Figure 1. Convergence tests on $\chi^{(3)}(\omega)$ with (a, b, c) the 10×10×10, 20×20×20, 30×30×30, and 40×40×40 k-points meshes and (d) with the number of empty states per atom, *i.e.*, 10, 14, and 18 empty states per atom.

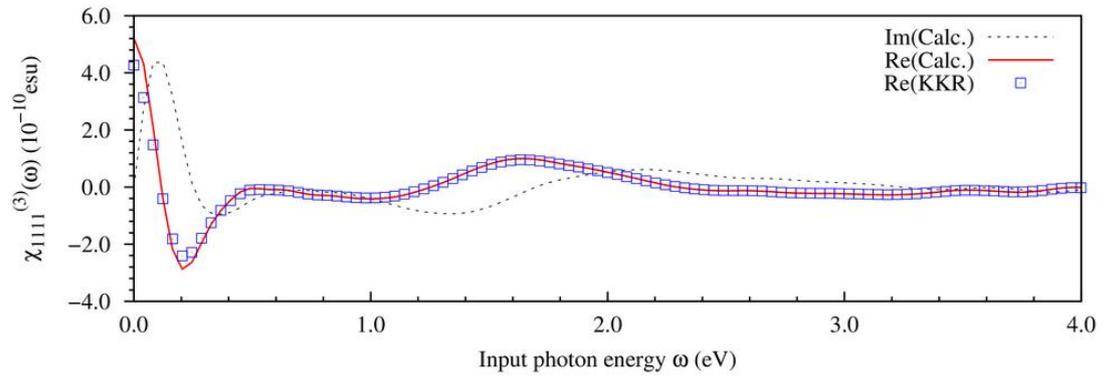

Figure 2. KKR test from imaginary (Im) to real (Re) part of $\chi^{(3)}(\omega)$. The dot and solid lines indicate the directly calculated results while the hollow square indicates the results based on the KKR calculation.

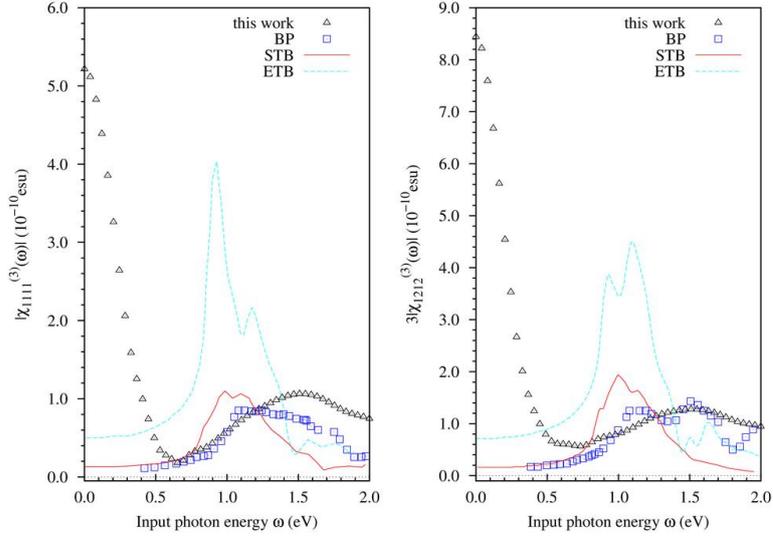

Figure 3. Dispersion of the two independent components of $\chi^{(3)}(\omega)$ based on our strategy, the tight-binding calculations with either semi-*ab initio* (STB) or empirical parameters (ETB), and *ab-initio* approach by means of the dynamical Berry phase (BP).

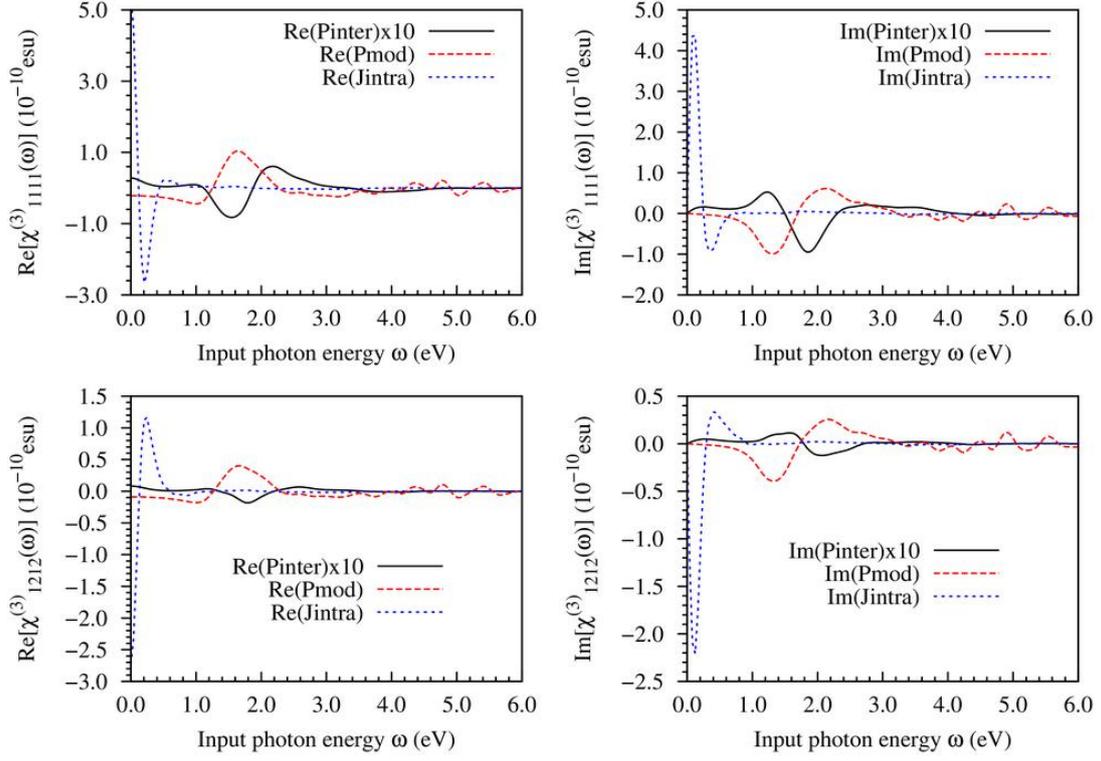

Figure 4. Dispersions of the pure interband (Pinter), a modulation of interband terms by intraband terms (Pmod), and intraband (Jinter) contributions of $\chi^{(3)}(\omega)$. P and J indicate $\mathbf{P}_\chi$ and $\mathbf{J}_\sigma$ components, respectively. For clarity, the pure interband contribution is increased by an order of magnitude (*i.e.*, ×10).

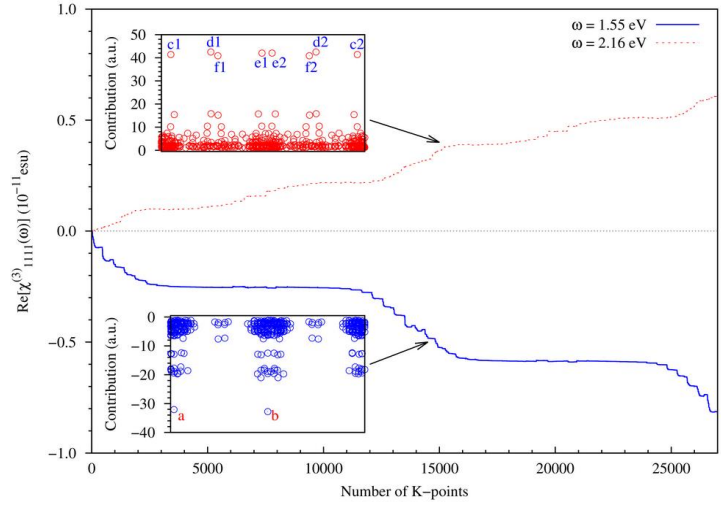

Figure 5. Traces of summation over all 27000 (30×30×30) k-points for the real part of $\chi^{(3),\text{Pinter}}_{1111}(\omega)$ with $\omega$ = 1.55 and 2.16 eV. Insets are the distribution of contribution (a.u.) per k-points in summation and for clarity only absolute contributions larger than 1.0 a.u. are shown.

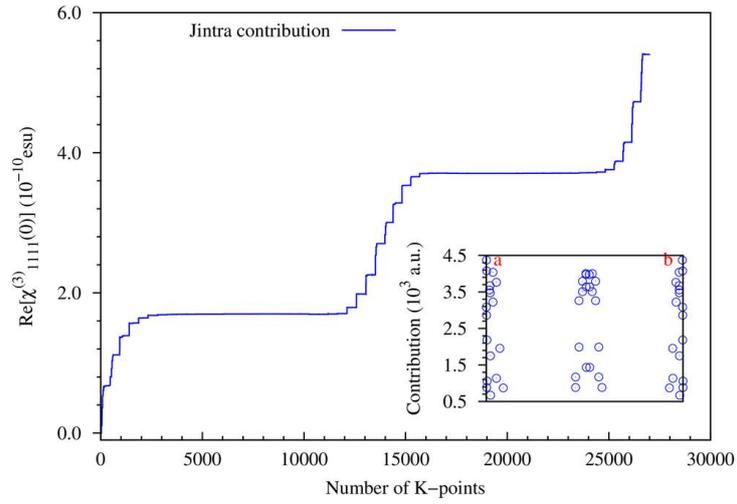

Figure 6. Traces of summation over all 27000 (30×30×30) k-points for the real part of $\chi^{(3),\text{Jintra}}_{1111}(0)$. The inset is the distribution of contribution (a.u.) per k-points in summation and for clarity only absolute contributions larger than 500 a.u. are shown.

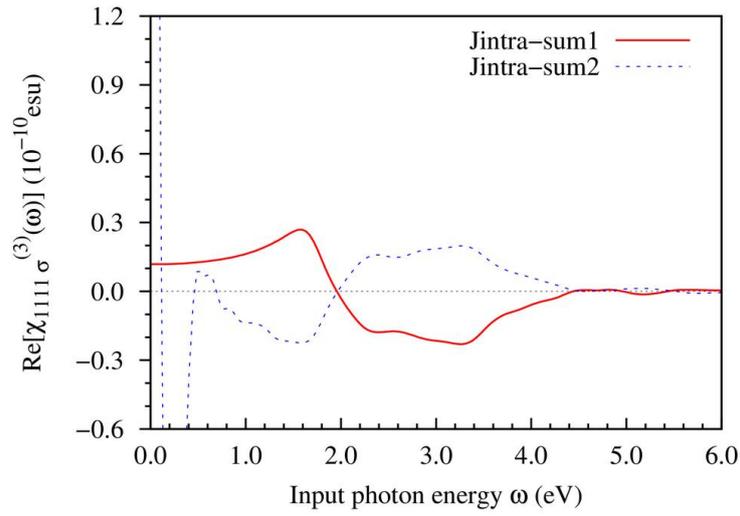

Figure 7. Dispersion behaviors of the first (Jintra-sum1) and second (Jintra-sum2) summations of $\chi^{(3),\text{Jintra}}_{1111}(\omega)$ (Eq. 2 in text).